\documentclass[10pt]{article}
\usepackage{graphics,epsfig}

\font\smallroman=cmr9

\font\bigroman=cmr12

\begin{document}
\title{Complementary Descriptions (PART I)\\ {\bigroman A Set of Ideas Regarding the Interpretation of Quantum Mechanics}}
\author{Christian de Ronde}
\date{}
\maketitle \centerline {Center Leo Apostel (CLEA), and}
\centerline {Foundations of the Exact Sciences (FUND),}
\centerline {Faculty of Science,} \centerline {Brussels Free
University} \centerline {Krijgskundestraat 33, 1160 Brussels,
Belgium.} \centerline {cderonde@vub.ac.be}

\begin{abstract}
\noindent Niels Bohr introduced the concept of complementarity in
order to give a general account of quantum mechanics, however he
stressed that the idea of complementarity is related to the
general difficulty in the formation of human ideas, inherent in
the distinction between subject and object. The complementary
descriptions approach is a framework for the interpretation of
quantum mechanics, more specifically, it focuses in the
development of the idea of {\it complementarity} and the concept
of {\it potentiality} in the orthodox quantum formulation. In PART
I of this article, we analyze the ideas of Bohr and present the
principle of complementary description which takes into account
Einstein's ontological position. We argue, in PART II, that this
development allows a better understanding of some of the
paradigmatic interpretational problems in quantum mechanics, such
as the measurement problem and the quantum to classical limit. We
conclude that one should further develop complementarity in order
to elaborate a consistent worldview.
\end{abstract}\bigskip
\noindent

\section*{Introduction}

I believe that scientific thought is intrinsically related with
desire and passion. Desire to create new ways to ``understand"
reality, passion to follow the path. This article is a compendium
of ideas and intuitions regarding the interpretation of quantum
mechanics, it should work as a ``road sign". It is certainly {\it
not} a closed system, but rather the opposite, a framework that
can allow for `new ideas' regarding the interpretation of the
quantum theory. I worn the reader he will not find closed
statements nor conclusions, rather, he might find some thoughts
which he is free to judge as interesting or not. Many of the
statements in this article are still to be worked out, my only aim
is to produce a suitable atmosphere where these ideas can be
developed, this is what I call the complementary descriptions
approach.

In this first part of the paper\footnote{The second part of this
paper, \cite{deRondeCDII}, is a more technical development of the
ideas presented in this article.}, I would like to present the
principle of {\it complementary descriptions} as mentioned in
\cite{deRonde03} and \cite{deRonde04}. To some extent I take this
principle to be an extension of (my reading of) Niels Bohr and
Wolfgang Pauli. It constitutes a general framework and a
philosophical worldview. We have learned many new insights in
quantum mechanics since the early discussions of Bohr and
Einstein. Thus, I believe the philosophy of complementarity first
introduced by Bohr \cite{Bohr28} in his `Como lecture' in 1927
should be also further developed in order to interpret and
understand the problems still present in the quantum theory.

The article is organized in four main sections. In section 1, I
present the principle of complementary descriptions. In section 2,
I give some general definitions of what is meant by a {\it
description}, a {\it perspective}, and a {\it context}. Section 3
deals with the concept of `complementarity'. Firstly, I analyze
Bohr's concept of complementarity, secondly, Pauli's view on
complementarity; thirdly, I give an account of the concept of
complementarity within the complementary descriptions approach. In
section 4, I study the idea of convergence in science. Finally, I
present the conclusions.

\section{Complementary Descriptions}

During the last centuries, there has been a tough struggle between
two main philosophical views of the world. On one side stands the
{\it empiricist view of the world}: the world is only a
conjunction of perceptions and phenomena\footnote{This position
has two main views, an ontological one, which was presented by
Berkley and Locke; and an epistemological view, which is defended
by Van Fraassen.}. There is no sense in looking for an external
reality, outside of my own perceptions. The only thing I have is
experience. In the 20th century, instrumentalism inherited many
tenets of the traditional empiricism: according to the {\it
instrumentalistic view}, a physical theory should link measurable
phenomena without being concerned with what happens to nature when
it is not measured. The theory does not seek to describe what
nature is, but rather it is a mere instrument of prediction, and
only in that sense it is an instrument of knowledge. Following
Mach, the function of a physical theory is to provide a simple and
economical synopsis of observable phenomena. It can be used to
make predictions of future phenomena, but it must not be thought
as providing a description of the world behind the phenomena. The
theory is an instrument and not a description. Opposite to this
position, we have the {\it realistic view of the world}: this view
stands for the independent and objective reality of the world. The
article you have at this moment in your hands, would be in front
of you, even if you would not be reading it, even if you would
close your eyes, even if you would not be able to experience it in
any way. There is an external world whose existence does not
depend on the presence of observers. Consciousness is a contingent
creation of nature and it does not determine the physical
existence of the world. Thus, a physical theory must be regarded
as the enterprise of discovering and describing the basic
structure of the world behind the phenomena. This point of view
has been the guide for the physical discovery of the world since
Tales tried to explain, without the use of mythical tales, how the
world was one of many rocks in the universe.

In quantum mechanics the realist-antirealist debate was exposed by
two of the most important figures of the quantum revolution:
Albert Einstein and Niels Bohr. The first regarded science from an
ontological basis and stressed that: {\it ``it is the purpose of
theoretical physics to achieve understanding of physical reality
which exists independently of the observer, and for which the
distinction between ``direct observable" and ``not directly
observable" has no ontological significance; this aim furnishes
the physicist at least part of the motivation for his work; but
the only decisive factor for the question whether or not to accept
a particular physical theory is its empirical
success."}\footnote{Einstein quoted from \cite{Dieks88a}, p.175.}
On the other hand, Niels Bohr regarded science from an
epistemological point of view. According to his long time
assistant, Aage Petersen, Bohr once declared when asked whether
the quantum world could be considered as somehow mirroring an
underlying quantum reality:

{\smallroman
\begin{quotation}
``There is no quantum world. There is only an abstract quantum
physical description. It is wrong to think that the task of
physics is to find out how nature is. Physics concerns what we can
say about nature" N. Bohr quoted from A. Petersen
(\cite{Petersen63}, p.8) \end{quotation}}

I would like to avoid endless discussions about what Bohr `really'
meant to say, or if he did actually say something like
this\footnote{In a talk in Buenos Aires David Mermin (September
29, 2003) presented this same quotation after which he mentioned
that some physicists (who had known Bohr personally) agreed and
disagreed about whether Bohr could have said something like
this.}. Rather, I would like to present here my own position and
extend this as follows: {\it There is no quantum world, nor is
there a classical world. There is only an abstract quantum
description as there is an abstract classical description.} With
respect to the latter I want to take a middle path between the
epistemological restrictions imposed by our means of expression
and the ontological commitment which Einstein so vividly defended.
{\it Physics not only concerns what we can say about nature but
also about what nature {\bf is}}.

Methodological reductionism states that the understanding of a
complex system is best sought at the level of the structure and
behavior of its component parts. Hans Primas (\cite{Primas83},
p.308) wrote with respect to this matter: {\it``In spite of the
obvious plurality of scientific explanations on various levels of
descriptions, there still exists a bias toward theoretical monism.
Neglecting the possibility to view nature from different
perspectives and ignoring the fact that the decomposition of
nature into parts is not God-given, traditional reductionism
treats the various theories and models as, to be sure,
incompletely articulated but ultimately reducible to an all
embracing fundamental theory."} Opposite to methodological
reductionism stands methodological holism, which states that the
understanding of a complex system is best sought at the level of
principles governing the behavior of the whole system, and not at
the level of the structure and behavior of its component parts
(\cite{Healey99}, p.2). Half way between these positions, stands
{\it complementarity}\footnote{I should clearly stress that
however, I do not try to present the complementary descriptions
approach as a closed method; rather, it supports Feyerabend's
\cite{Feyerabend93} critic: {\it ``The idea that science can, and
should, be run according to fixed and universal rules, is both
unrealistic and pernicious.[...] All methodologies have their
limitations and the only `rule' that survives is `anything
goes'."} I will develop the meaning of complementarity within this
approach in Sect. 3.3.}. Niels Bohr introduced the concept of
complementarity in order to describe situations in which two
different conditions of observation yield conclusions that are
conceptually incompatible. However, Bohr \cite{Bohr28} made it
clear that complementarity is not restricted to quantum physics.
He stressed that the idea of complementarity is related {\it ``to
the general difficulty in the formation of human ideas, inherent
in the distinction between subject and object"}. Hans Primas makes
a clear statement of the importance of such a concept:

{\smallroman
\begin{quotation}
``A new style in science began with quantum mechanics when Niels
Bohr initiated a spiritual renewal by introducing his concept of
complementarity. It turned out that complementarity is far more
important than quantum mechanics, it has led to a development of
science that encourages a holistic vision." H. Primas
(\cite{Primas83}, p.349)
\end{quotation}}

I want to develop complementarity in order to take into account
not only {\it complementary contexts} (phenomena in Bohr's
terminology) but also {\it complementary descriptions} such as the
{\it classical description} (which stresses the reductionistic
character of the being) and the {\it quantum description} (which
stresses the holistic character of the being). For this purpose I
will follow a middle path between ontology (a certain form of the
being) and epistemology (the theoretic preconditions of the
description which determine a certain access to the being). I take
{\it complementarity} as a philosophical term, connected with the
{\it ``limitations of our means of expression"}\footnote{See for
example von Weiszacker (\cite{vWeizsacker85}, p.282) or Heisenberg
(quoted in \cite{WheelerZurek83}, p.86) for a equivalent statement
of the concept of complementarity; also Bohr's quotations in
(\cite{Plotnitsky94}, pp.68, 72 and 121) and in
(\cite{WheelerZurek83}, p.88).}. In this sense special attention
should be paid to the observer as precondition of the physical
reasoning and not as object of the theory.

Immanuel Kant expressed the idea that the conditions of
possibility for knowledge are given by a definite group of
categories which together with the forms of sensible intuition
(space and time) constitute the object, however, we do not have
access to `the thing in itself' which remains forbidden to us:

{\smallroman
\begin{quotation}
``Not only are the drops of rain mere appearances, but even their
round shape, and even the space in which they fall, are nothing in
themselves, but merely modifications of fundamental forms of our
sensible intuition, and the transcendental object remains unknown
to us" I. Kant (\cite{Kant}, p.85)
\end{quotation}}

Kant critically analyzed the ``metaphysical foundation of science"
which was based in classical logic and classical mechanics; he
patterned the categories after the table of judgements and the
synthetic principles after Newton's laws of gravitation. All
genuine science requires a `pure part' which must be worked out
separately {\it ``so that we may strictly determine what reason
can accomplish by itself and where it begins to need the aid of
principles of experience"}\footnote{Kant, quoted from
\cite{Cassirer56}.}. I would like to analyze in the following
sections the problem of the limitations in the framework presented
by Kant: {\it``We know today why this pure part could not fulfill
the task that Kant put to it. It was too closely bound to a
specific form of science, which classical rationalism held to be
the plainly rational form."}\footnote{Furthermore Cassirer
(\cite{Cassirer56}, p.74) also expresses with respect to the
specific problem in quantum mechanics: {\it``When in the face of
the new factual material and new theoretical tasks which it was
facing, physics extended and transformed its conceptual apparatus,
it did not simultaneously give up its general character and
structure. It only made evident the fact that this structure is to
be thought of not as rigid but as dynamic, that its significance
and efficacy do not rest upon substantial rigidity established
once and for all but precisely upon its plasticity and
flexibility."}} Regarding this particular criticism of the program
developed by Kant, Wolfgang Pauli makes the following remark:

{\smallroman
\begin{quotation}
``We agree with P. Bernays in no longer regarding the special
ideas, which Kant calls synthetic judgements a priori, generally
as the pre-conditions of human understanding, but merely as the
special pre-conditions of the exact science (and mathematics) of
his age." W. Pauli (\cite{Pauli94}, p.126)
\end{quotation}}

To go beyond Kant we must take new concepts into play and not only
regard fixed categories in the synthesis. The first step is thus,
to develop these new concepts. I don't take for granted the
justification over the condition of objectivity, however in order
to apply such a procedure we must first have the new concepts at
hand. In our schema there is no complete cut between `the thing in
itself' and that which we regard as the real. There is access to
the {\it real} as an expression of a certain description exposed
to the experimental observation. Reality comes together from the
synthesis of the {\it creative} element of our description and the
{\it discovery} character of the empirical data. `The thing in
itself' is not the description nor the experimental observation,
it is not accessible as a whole and in this sense it remains {\it
veiled}\footnote{This term was introduced by Bernard D'Espagnat
\cite{D'Espagnat95}.}.

Ontology is the study of the {\it being}, it is not {\it the real}
itself but rather its study, thus, always of a specific form, a
restricted aspect of the being. I will take a position which will
take into account the distinction of different complementary
descriptions\footnote{This is in close analogy to the internalist
position developed by Hilary Putnam \cite{Putnam81} and also by
Hans Primas \cite{Primas83}, \cite{Primas87}, \cite{Primas94}.}.
Within this approach ontology only arises from the synthesis
between our conceptual scheme and the noumenal realm\footnote{The
noumenon is taken by Kant to be that which is not accessible by
the senses, which remains completely unknown to us: {\it``das Ding
an sich"}. As we expressed above we will not follow this cut but
rather state there is a certain access to the {\it real}.} as
exposed in the experimental observation. Different conceptual
schemes (descriptions) define different ontologies through their
`imprint' in the experimental observation. It is important to
stress the idea that different descriptions define different, in
prinicple incompatible, ontologies. Ontology is description
dependent, so ``objective" does not mean independent of the
subject, but rather the opposite, it results from applying our
different frameworks (descriptions); the more possibilities and
abstract conceptual schemes one is able to apply to
describe/create reality the more ``objective" one gets\footnote{I
am grateful to Pablo Vitalich for discussing this idea with me.}.
Within the position taken by the complementary descriptions
approach different levels of discourse can define equally
``objective" ontologies, each of which makes possible the creation
of a certain experience. I want to stress the importance of
knowing from which descriptive level one is talking from, because
it might happen that the same sentence can have only significance
in a certain level while it is meaningless in a different one
(\cite{deRonde03}, Sect. 9.3). I will argue that only together,
different complementary descriptions allow a resolution of the
quantum paradoxes and a better `understanding' of the quantum
theory of measurement.

Bohr explains that for the description of certain atomic phenomena
we need a particle picture while for others we need the wave
picture. Using both pictures simultaneously leads to
contradictions; it is the concept of {\it phenomenon} which allows
to avoid such controversies. According to Bohr one never needs
{\it complementary views} in the same phenomenon and objectivity
is regained.

{\smallroman
\begin{quotation}
{``On the lines of objective description [I advocate using] the
word {\it phenomenon} to refer only to observations under
circumstances whose description includes an account of the whole
experimental arrangement. In such terminology, the observational
problem in quantum physics is deprived of any special intricacy
and we are, moreover, directly reminded that every atomic
phenomenon is closed in the sense that its observation is based on
registrations obtained by means of suitable amplification devices
with irreversible functioning such as, for example, permanent
marks on a photographic plate, caused by the penetration of
electrons into the emulsion. In this connection, it is important
to realize that the quantum-mechanical formalism permits well
defined applications referring only to such closed phenomena."} N.
Bohr (\cite{Bohr58}, p.73, quoted from \cite{WheelerZurek83}, p.3)
\end{quotation}}

The definition of phenomenon relied for Bohr in the use of {\it
classical language}. We want to extend this definition by allowing
the possibility of expressing a phenomenon with non-classical
descriptions, the concepts of which are still to be
developed\footnote{This goes clearly against the conception of
Bohr (\cite{WheelerZurek83}, p.7) who stated that: {\it ``it would
be a misconception to believe that the difficulties of the atomic
theory may be evaded by eventually replacing the concepts of
classical physics by new conceptual forms."}}. In our approach
there are no `naked facts'. This goes against a {\it naive
empiricist} position which would identify reality with
experience\footnote{This positivist position renders objectivity
to facts, the limitations of such position will be discussed in
\cite{deRondeCDII}, Sect. 1.5.}. A phenomenon is thus the
synthesis between the description (which determine the conditions
of possibility to access a certain aspect of the being) and
experimental observation (the noumena as exposed by the
description). Quantum theory should be developed in order to
extreme it's own concepts, we need to help quantum mechanics find
its own interpretation. This can be pursued by a careful analysis
of its historical development, taking special attention to the
ideas developed by the founding fathers of the theory.

\section{Descriptions, Perspectives and Contexts}

I consider {\bf descriptions} to be a general framework in which
concepts are related. Descriptions express the precondition to
access a certain character of reality. Descriptions are closed
systems, they interrelate but they are not necessarily
complete\footnote{I will discuss in more detail the idea of
completeness and incompleteness in Sect. 3.3}. A description is
defined by a specific set of concepts, this definition precludes,
at a later stage, the possibility of applying different
(incompatible) concepts. Quantum mechanics is a description as
well as classical mechanics or relativity theory are, each of them
relating concepts which are not necessarily compatible with the
ones present in a different description. Objects are already part
of a {\it particular} description with strong presuppositions. As
a matter fact, we will argue that the quantum description is
incompatible with the preconceptions needed to define an object in
the classical sense (I will go more deeply into this matter in
\cite{deRondeCDII}, Sect. 1.1). This goes against a {\it naive
realist} position which would identify reality with the objects
around us. Articles, tables and Stern-Gerlach apparatus are not
fundamental blocks of reality, they are conceptual creations we
are forced to make in order to approach a certain character of
reality\footnote{In this sense my position is close to that of a
critical realism.}. The main characterization of a description is
then by means of the concepts applied which deal with the
irreducible limitation and incompleteness in the means of
expression that must be acknowledged in every description. For
example, the concepts of mass, space and time in classical
mechanics are not the same concepts of mass, space and time in
general relativity theory; we are dealing with different concepts
which have developed historically and are determined by different
descriptions. The fact that we, as physicists, use the same words
is quite misleading\footnote{On this respect see also the
quotation of Wittgenstein on p.10 of this article.}. It is
relevant to state at this point that I do not take physics to be
the only possible description. Philosophical thought, psychology,
biology or economy can be viewed as general descriptions which, in
the same way as physics, try to grasp certain `knowledge', certain
`understanding'. The idea of encapsulating this knowledge into
different `boxes' should not be pushed to the extent it is done
today.

I take a {\bf perspective} to be a {\it faculty}. It expresses the
{\it potentiality} of an action which makes possible a definite
consistent view within a certain description. A perspective is the
condition of possibility for a definite representation to take
place, it deals with the choice between mutually incompatible {\it
contexts}. The importance of defining this level of description
(which can be regarded as superfluous in the classical realm) will
become clear in \cite{deRondeCDII}.

The {\bf context} is a {\it definite representation} of the
entity. Contexts are determined by the conceptual scheme of the
description; i.e. they are description dependent. This leaves open
the possibility of representing experiments with `new'
non-classical concepts.

The distinction between {\it perspective} and {\it context} was
introduced in \cite{deRonde03} in order to distinguish between the
{\it holistic properties} and the {\it reductionistic properties}
in the Bene-Dieks perspectival interpretation \cite{BeneDieks02}
(this will be further analyzed in \cite{deRondeCDII}; Sect. 1.4).
In special relativity theory a {\it context} is given by a
definite inertial frame of reference. There is no necessity of
defining the perspective because the invariance principle allows
us to think of these contexts as existing in {\it actuality}, as
events which pertain to {\it physical reality}. In quantum
mechanics a {\it context} is given by a definite experimental set
up; i.e. a Complete Set of Commuting Observables (C.S.C.O.) which
is defined equivalently by a Schr\"odinger wave function in a {\it
definite basis/representation}. It is only at this stage that we
can talk of a quantum-entity (a set of mutually compatible
properties\footnote{We refer to the meaning of compatibility in
\cite{AertsdeRonde}.}). Different contexts, due to the commutation
relations as expressed in the standard formulation, can be
mutually incompatible; thus, the {\it quantum contexts} cannot be
thought as existing in {\it actuality}. In classical mechanics
(and relativity theory) this problem does not arise because of the
relation of contexts through the Galilean (and Lorentz)
transformations.

The main difference between quantum mechanics and the rest of the
theories created by man is that the quantum wave function
expresses {\it explicitly} a level in the description of nature
which has been neglected from a mechanistic idea of a
clock-type-world. It presents us with the concept of {\it choice}
within knowledge itself. This character is expressed by the {\it
perspective} which lies in the {\it indeterminate level} (that
which has to be chosen) while the {\it context}\footnote{The {\it
perspective} can be regarded in this sense as the wave function
without any definite basis (or factorization in a scheme with a
preferred basis), it is an expression of the possible contexts.
The {\it context} can be seen as the explicit election of the
`representation'; i.e. the basis or factorization. In quantum
mechanics this choice determines explicitly the entity under study
(this character will be further discussed in \cite{deRondeCDII},
Sect. 1.2 and 1.3).} lies in the {\it determined level} (that
which has been chosen). However, in the quantum context there is
still certain indetermination regarding the properties in the
sense that a superposition expresses still the potential, and in
this sense {\it is} and {\it is not}. Concepts are lacking in the
development of these levels. I will introduce the concept of {\it
ontological potentiality} as an attempt to escape the limits of
classical thought (this will be analyzed in more detail in Sect.
1.2 of \cite{deRondeCDII}).

\section{The Concept of `Complementarity'}

In this section I will go through several views regarding the
concept of complementarity. Firstly, I will give an account of
complementarity as introduced by Bohr, secondly, I will present
the ideas of Wolfgang Pauli with respect to complementarity,
finally, I will present the meaning of complementarity within our
approach.

\subsection{Bohr's Concept of `Complementarity'}

The incompatibility of different views is expressed by the
commutativity relations of Heisenberg which show that the {\it
position-context} and the {\it momentum-context} are mutually
incompatible\footnote{This is the expression that experimental
arrangements which show particle properties preclude the
possibility of showing wave properties and vice versa.}:

\begin{equation}
{\lbrack x,p]\neq 0}
\end{equation}

The great achievement of Bohr was to bring together these mutually
excluding {\it contexts} by means of the concept of
complementarity. As John Hendry (\cite{Hendry84} p.119) expresses:
{\it ``Of all those actively involved in the search for a new
quantum mechanics in the 1920s, Bohr was at once the most radical
and the most conservative. He had been initially responsible for
the idea that classical mechanics and kinematical concepts were
incapable of describing quantum phenomena, and he had continued to
believe this throughout. But he had also held fast to the belief
that these concepts, and especially those of the classical wave
theory of light could not be replaced."} Heisenberg had always
emphasized the discreteness of quantum theory. This conception
brought him to the uncertainty relations in February 1927 when
Bohr left Copenhagen for a holiday. At his return, the priority of
the particle picture over the wave particle in the scheme
presented by Heisenberg was not taken by Bohr very
enthusiastically and discussions followed in which Pauli had to
defend his young fellow. The objections raised by Bohr, which
would later be expressed in \cite{Bohr28}, can be tracked to an
``Addition in Proof" at the end of Heisenberg's paper:

{\smallroman
\begin{quotation}
``After the conclusion of the forgoing paper, more recent
investigations of Bohr have led to a point of view which permits
an essential deepening and sharpening of the analysis of
quantum-mechanical correlations attempted in this work. In this
connection Bohr has brought to my attention that I have overlooked
essential points in the course of several discussions in this
paper. Above all, the uncertainty in our observation does not
arise exclusively from the occurrence of discontinuities, but is
tied directly to the demand that we ascribe equal validity to the
quite different experiments which show up in corpuscular theory in
the one hand, and in the wave theory in the other hand. [...] I
owe great thanks to Professor Bohr for sharing with me at an early
stage the results of these more recent investigations of his-to
appear soon in a paper on the conceptual structure of quantum
theory- and for discussing them with me." W. Heisenberg
(\cite{Heisenberg27} quoted from \cite{WheelerZurek83}, p.83)
\end{quotation}}

One can see the importance (and pressure) of Bohr's figure upon
the young Heisenberg in one of the foundational articles of
quantum mechanics. Bohr's own search of a consistent
interpretation was later explained by Leon Rosenfeld:

{\smallroman
\begin{quotation}
``Bohr wanted to pursue the epistemological analysis one step
further, and in particular to understand the logical nature of the
mutual exclusion of the aspects opposed in the particle-wave
dualism. From this point of view the indeterminacy relations
appear in a new light. [...] The indeterminacy relations are
therefore essential to ensure the consistency of the theory, by
assigning the limits within which the use of classical concepts
belonging to the two extreme pictures may be applied without
contradiction. For this novel logical relationship, which called
in Bohr's mind echoes of his philosophical meditations over the
duality of our mental activity, he proposed the name
``complementarity", conscious that he was here breaking new ground
in epistemology." L. Rosenfeld (quoted from \cite{WheelerZurek83},
p.59)\end{quotation}}

As we already know, there are as many definitions of
complementarity as physicists attempting to define it\footnote{See
for example the analysis of complementarity and Bohr's philosophy
in \cite{Howard94} and \cite{Saunders05}.}. What Bohr expressed by
`complementarity' remains foggy to me. I was to some extent
released when I found out other people shared my problem. The most
weird example I found was the case of Carl Friedrich von
Weizs\"acker who wrote an article named: {\it ``Komplementarit\"at
und Natuurwissenschaft"} \cite{vWeizsacker55} for the 70th
birthday of Niels Bohr. In this article he explained the concept
of complementarity in two different forms, namely, {\it parallel
complementarity} and {\it circular complementarity}\footnote{{\it
Parallel complementarity} is defined by von Weizs\"acker as the
complementary relation which takes place between, for example,
{\it position} and {\it momentum} or {\it particles} and {\it
waves}; this is very close to what I call {\it complementarity
between contexts}. On the other hand {\it circular
complementarity} was defined by von Weizs\"acker as
complementarity between {\it classical concepts} and the
description given by the {\it Schr\"odinger wave function}; this
is very close to what I call {\it complementarity between the
quantum and the classical descriptions}.}, the second of which was
attributed to Bohr himself. The article ends with a rectification
of von Weizs\"acker in which he explains that he received a letter
from Bohr (\cite{vWeizsacker74}, p.338) expressing that
complementarity can be only defined with respect to
phenomena\footnote{A clear expression of this insight in Bohr's
concept of complementarity can be found in (\cite{WheelerZurek83},
p.3): ``{\it Complementarity: any given application of classical
concepts precludes the simultaneous use of other classical
concepts which in a different connection are equally necessary for
the elucidation of the phenomena."}}, and as the Schr\"odinger
wave equation is just an {\it abstract magnitude of calculus} and
it does not designate in itself any phenomena, such circular
complementarity is by no means possible and only parallel
complementarity should be taken into account\footnote{See also
Jammer's remarks on this episode in \cite{Jammer74}.}. Even von
Weizs\"acker, who was an active player in the discussions that
took place during the development of quantum theory, all together
with Bohr himself, Pauli and Heisenberg, had a misleading idea of
what Bohr meant with complementarity.

Only a few thinkers are able to develop themselves trough history,
to discuss with non-contemporary people. I think the common
attitude nowadays is to read Bohr and then try to convince others
that if Bohr would not be dead he would sit at the same side at
the discussion table. I believe that on the contrary, Bohr himself
is still quite alive and if one tries enough, it even is possible
to discuss with him.

\subsection{Pauli's Concept of `Complementarity'}

Wolfgang Pauli was one of the most brilliant and creative minds of
the last century, he was called by his colleagues ``the
consciousness of science." He discussed deeply the implications of
complementarity with Bohr himself and he saw in its philosophy a
more general way of approaching science as is expressed in the
following quotation:

{\smallroman
\begin{quotation}
``...in the hope of furthering by this small contribution those
major efforts which have the general aim once more bringing into
closer contact the various partial disciplines into which our
intellectual life (Geistigkeit) has fallen apart. The splitting
off of the exact sciences and mathematics as independent partial
disciplines from an originally unified but pre-scientific natural
philosophy, which began in the 17th century, was of course a
necessary condition for the subsequent intellectual development of
the western world (Abendland). At the present time, however, the
conditions for a renewed understanding between physicists and
philosophers on the epistemological foundations of the scientific
description of nature seem to be satisfied. As a result of the
development of atomistics and quantum theory since 1910 physics
has gradually been compelled to abandon its proud claim that it
can, in principle, understand the whole universe. All physicists
who accept the development that reached a provisional conclusion
in 1927 in the systematic construction of the mathematical
formalism of wave mechanics, must admit that while at present we
have exact sciences, we no longer have a scientific picture of the
universe (Weltbild). It is just this circumstance that may contain
in itself, as a corrective to the earlier one sided view, the germ
of progress towards a unified total world-picture, of which the
exact sciences are only part. In this I would like to see the more
general significance of the idea of complementarity, an idea that
has grown out of the soil of physics, as a result of the work of
the Danish physicist {\it Niels Bohr}." W. Pauli (\cite{Pauli94},
p.36)\end{quotation}}

I think Pauli could see the deeper meaning of complementarity even
more clear than Bohr himself, he extended the principle to
knowledge itself, seeking for a new type of science. For example,
he repeatedly stressed the idea of complementarity between physics
and psychology, a problem which he called ``the most important
problem of our time".

{\smallroman
\begin{quotation}
``It actually seems to me that in the {\it complementarity in
physics}, with its resolution of the wave-particle opposites,
there is a sort of {\it role model or example of that other, more
comprehensive coniunctio}. For the smaller {\it coniunctio} in the
context of physics, completely unintentionally on the part of its
discoverers, has certain characteristics that can also be used to
resolve the other pairs of opposites listed on p. 3.
[...psychology-physics...]" W. Pauli (\cite{PauliJung}, p.91)
\end{quotation}}

Pauli was aware of the importance of a true development of the
concept of reality, a concept which is at the main core of the
ontological problems raised by quantum mechanics. He writes:

{\smallroman
\begin{quotation}
``When the layman says ``reality" he usually thinks that he is
speaking about something which is self-evidently known; while to
me it appears to be specifically the most important and extremely
difficult task of our time to work on the elaboration of a new
idea of reality." W. Pauli (quoted from \cite{Laurikainen98},
p.193)
\end{quotation}}

The quotations I have chosen express a small part of Pauli's
thought. The path created by Pauli's deep thinking is the one I
will try follow in the sections to come.

\subsection{`Complementarity' within the Complementary Descriptions Approach}

The idea that reality can be captured by an objective system has
been to great extent the motor of knowledge in its different
forms. The acknowledgment of the impossibilities for such a
project, should not be regarded as a defeat but rather as a new
insight of reality itself. In this same sense I don't take
complementarity as describing an encompassing whole, on the
contrary, I think complementarity obliges us to face the
incompleteness and paradoxical character of reality. {\it
``Complementarity is no system, no doctrine with ready-made
precepts. There is no via regia to it; no formal definition of it
can be found in Bohr's writings, and this worries many people.
[...] Bohr was content to teach by example. He often evoked the
thinkers of the past who had intuitively recognized dialectical
aspects of existence and endeavored to give them poetical or
philosophical expression."}\footnote{Rosenfeld quoted from
(\cite{WheelerZurek83}, p.85).} Complementarity goes with paradox,
it allows us to stress the limits of knowledge and at the same
time it presents us with the incommensurability of reality.
Descriptions, perspectives, contexts and concepts are then taken
{\it as complementary} in this same sense.

With the complementary descriptions approach I try to find a {\it
middle path} between ontology and epistemology between realism and
empiricism. The main difficulty of this approach is to stand in
between, to not be dragged by any specific description, each of
which should be regarded only as a `partial description', and
complementary to a different one.

I regard quantum mechanics, on the one hand, as a {\it complete}
theory; in the sense of Pauli (\cite{Pauli94}, p.96): {\it ``that
no new laws can be added to the system of natural laws of the
domain connected without partly altering the content of those
already contained in it."}\footnote{This is the same sense in
which Heisenberg regarded quantum mechanics as a {\it complete
theory} and which many times has been badly characterized as
meaning that quantum mechanics is {\it the} fundamental theory of
nature. As expressed in \cite{vWeizsacker85} by von Weizsacker:
{\it ``For Heisenberg, ``closed" was not to identical with
``final", but the sequence of closed theories rather hinted at the
idea of physics as an open-ended enterprise."} See also
\cite{Bokulich04}.} On the other hand, the EPR argument, as Aerts
\cite{Aerts85} correctly points out, is an {\it ad absurdum} proof
of the {\it incompleteness} of the quantum theory. In what sense
quantum mechanics is incomplete, I argue, relates to the
complementarity of the classical and quantum concepts involved.
Every description is incomplete in the following: it has to choose
a certain structure, a certain group of concepts which define it.
In their turn these concepts expose the limit within which the
description is `useful'; but a limit is always in itself a
definition of something which lies `outside'. {\it``Insofar as it
is measured against and prohibits the classical models of
completeness, complementary entails incompleteness --structural,
irreducible incompleteness."}\footnote{Plotnitsky quoted from
(\cite{Plotnitsky94}, p.128).} This incompleteness deals
explicitly with the complementarity of descriptions, every
description expressing in itself a limit in our means of
expression and understanding. The incompleteness of quantum
mechanics presents us with its limit; for example with the
impossibility of describing separated entities (as it is
presupposed in classical mechanics/relativity theory). One should
be aware that in this sense classical mechanics is also incomplete
as it cannot by any means take into account the `non-separability'
present in the quantum formulation \cite{Schrodinger35}.

According to the received view our language presupposes
individuality, it is an expression of the metaphysical choice
inherited from Aristotle, Plato and the parmenidean `One'. These
presuppositions expressed by classical logic go together with
classical mechanics. Quantum mechanics, on the other hand is
closer to the heraclitean `Many'\footnote{See for example David
Finkelstein's article: ``All is Flux" \cite{Finkelstein87} and
\cite{VerelstCoecke99} for a related analysis.}. Classical and
quantum mechanics are definite descriptions with definite
concepts. Both theories approach reality, the result of which is
to a large extent, partially determined by the description itself:
they are mutually incompatible, and at the same time they are
expressions of the preconceptions involved presenting
complementary views of reality. It is in this general sense that
{\it I take the quantum mechanical description to be complementary
to the classical description}.

As Folse (\cite{Folse87}, p.163) argues: {\it``Bohr persistently
evades any direct engagement with the question of ``reality"."} I
will not argue for or against the `realist-antirealist' position
of which a lot of bibliography can be found\footnote{See for
example \cite{Folse85}, \cite{Howard89} and references therein.}
but rather, as I expressed above, try to move towards a new
conception of reality (of which this article might be considered
as a micro step). Reality should not be a pre-established concept
nor a prejudice in observing and relating empirical data, but
rather a goal concept which should be transformed and developed.
We should not expect reality to be... {\it as we would like it to
be}; we must constantly revise the conceptual framework with which
such a {\it description} is expressed. Following the main idea,
which led Einstein to the special theory of relativity, we should
not conclude experiments from reality; however, the opposite
should be neither pursued; Carlo Rovelli (\cite{Rovelli96}, p.2)
proposed the following: {\it``I have a methodological suggestion
for the problem of the interpretation of quantum mechanics:
Finding the set of physical facts from which the quantum
mechanics's formalism can be derived."} The problem is: there are
no {\it facts} without a {\it description!}\footnote{As Einstein
himself pointed out to Heisenberg: {\it ``It is only the theory
which decides what can be observed.}.} Experimental observation
and ontology are intricately related without supremacy of one over
the other, both being the reflection of the former like two
mirrors with nothing in between.

It must be clear at this point that I do not mean to take as a
standpoint, like Bohr did (\cite{WheelerZurek83}, p.7), the
importance of classical (physical) concepts in the definition of
{\it experience}. Bohr relied in a concept of complementarity
which was a consistent explanation of these {\it phenomena}. My
approach is a development of the concept of complementarity
stressing the importance of {\it descriptions} which make possible
the preconditions of experimental observation encouraging the
creation and development of {\it new concepts} within physics.
These descriptions develop a plurality of ontologies which capture
different forms of the being. To a great extent every new theory
that has been developed, from Aristotelian mechanics to general
relativity, has been grounded in new concepts. {\it The physicist
should be a creator of physical concepts}. Concepts which, within
a theory, make possible to grasp certain character of nature. This
however should not be regarded as some kind of solipsism, {\it it
is not only the description shaping reality but also reality
hitting our descriptions.} It is through this interaction, namely,
our descriptions and the experimental observation that we create
and discover a certain character of the being. It is in this way
that we can develop that which we consider to be {\it reality}.

My standpoint is that experience is defined by description, {\it
and vice versa}, description is defined by experience, they
intricate themselves with no preponderance of the one over the
other. In order to regain objectivity, we must acknowledge that
the classical description has no supremacy over different
descriptions, that it is not given to us {\it a priori}, and that
it develops through the different descriptions with which we
choose to express ourselves. The phrase of Bohr (quoted from
\cite{WheelerZurek83}, p.7) stating that {\it ``... the
unambiguous interpretation of any measurement must be essentially
framed in terms of the classical physical theories, and we may say
that in this sense the language of Newton and Maxwell will remain
the language of physics for all time."} gives way to a quite
strong conclusion, this is, that we have come to the limits in the
pre-conditions of human understanding. By having to confront
Einstein's realism and ontological position Bohr was dragged to
the other extreme, namely, an empiricist and epistemological
position. I hope to go back to the {\it middle path}, just in
between experience and description, between creation and
discovery\footnote{A related position, namely, the creation
discovery view was introduced by Diederik Aerts in
\cite{Aerts88}.}.

\section{Convergence of Descriptions?}

There is a quite tacit assumption which goes against the ideas I
have been presenting, namely, the idea that science is converging
towards the ultimate truth, the theory of everything, that our
knowledge increases with every article that is published. I think
the concept of what is understanding has been severely damaged by
a radical positivistic attitude science has taken in the last
centuries. I point this out because the idea that more production
is equal to more knowledge is at stake. One should, as a scientist
wonder over the very deep meaning of ``understanding"\footnote{See
for example the very interesting Chap. 6 of \cite{Heisenberg72}
about the concept of `understanding' in a discussion between
Heisenberg and Pauli.}. This maybe the core problem for many in
reaching the concept of complementarity, which tackles our tacit
presuppositions \cite{Folse87} in the traditional positivist
epistemological framework; i.e. that a theory provides knowledge
about an object if and only if it justifies making {\it true
descriptive statements} predicating properties of some substantial
entity. This idea rests on imagining that science has reached the
{\it a priori} conditions of human understanding itself while it
has in fact succeeded only in setting up the {\it a priori}
conditions of the systems of mathematics and the exact sciences of
a particular epoch (\cite{Pauli94}, p.95).

The idea of a convergent reality presupposes the idea that one can
reduce concepts of one theory to the next; i.e. that there is a
fundamental theory which can reach {\it the} fundamental `concepts
of nature'. Richard Feynman is a proponent of such view, in his
BBC television lectures he argued:

{\smallroman
\begin{quotation}
``The age in which we live is the age in which we are discovering
the fundamental laws of nature, and that day will never come
again. It is very exciting, it is marvellous, but this excitement
will have to go." R. Feynman (quoted from \cite{Primas83}, p.347).
\end{quotation}}

Reductionism goes together with convergence. In this sense,
classical mechanics is worse than relativity theory, because, the
last is able to see the concepts of the first as a limit, and at
the same time it produces new insights. Contrary to this position,
it is quite clear that when one studies the problem in a deeper
way, one finds that such concepts are no limits, rather, they can
be found as approximations within certain very specific
conditions; however, when these conditions are extended to the
general frameworks from which the concepts acquire meaning
incompatibilities and inconsistencies appear as much as in between
the classical and quantum descriptions. In other words, I think
that trying to find a limit between relativity theory and
classical mechanics is to some extent equivalent in trying to find
a limit between physics and psychology. Although concepts like
space and time can be used in the physical framework as well as in
the Freudian theory of psychology, once we generalize the concepts
to the general framework of either description we find out the
concepts generalize as well in both directions making impossible
to retain the consistency presented in the beginning. Quantum
mechanics is full of these type of mistakes which appear in most
cases by using concepts and symbols which are not part of the {\it
quantum language} (which make no sense in it). Through mixing
symbols and words which pertain to different languages in many
cases we end up in weird paradoxes. In order to get closer to the
mystery one first needs to demystify and clarify the limits and
the usage of the different languages. Wittgenstein was certainly
aware of this important fact as he points out in his {\it
Tractatus} (\cite{Wittgenstein}, pp.18-19):

{\smallroman
\begin{quotation}
{\bf 3.323.}  In everyday language it very frequently happens that
the same word has different modes of signification--and so belongs
to different symbols--or that two words that have different modes
of signification are employed in propositions in what is
superficially the same way.

Thus the world `is' figures as the copula, as a sign for identity,
and as an expression for existence; `exist' figures as an
intransitive verb like `go', and `identical' as an adjective; we
speak of {\it something}, but also of {\it something's} happening.

(In the proposition `Green is green'--where the first word is the
propper name of a person and the last an adjective--these words do
not merely have different meanings: they are {\it different
symbols}.)

{\bf 3.324.}  In this way the most fundamental confusions are
easily produced (the whole philosophy is full of them).

{\bf 3.325.}  In order to avoid such errors we must make use of a
sign-language that excludes them by not using the same sign for
different symbols and by not using in a superficially similar way
signs that have different modes of signification: that is to say,
a sign-language that is governed by {\it logical} grammar--by
logical syntax.
\end{quotation}}

The idea that quantum mechanics is a fundamental theory of nature
(as describing the fundamental blocks of reality from which
everything else can be derived), even the idea that there might
exist a true story about the world\footnote{See for example
\cite{Weinberg93}.} goes completely against the spirit of what I
am proposing here. My approach goes together with Heisenberg's
conception of {\it closed theories} as a relation of tight
interconnected concepts, definitions and laws whereby a large
field of phenomena can be described \cite{Bokulich04}. Heisenberg
states that theories come up through ``intelectual jumps" with the
theories that have come before. In this point he confronted the
ideas of P.A.M. Dirac who states that theories are continuously
revised and straightened \cite{Bokulich04}. My approach goes just
in between; the composite (discrete and continuous) development
makes possible to take into account both, holistic and
reductionistic conceptions of science; this is where
complementarity remains a fundamental position in order to find
new paths\footnote{I am grateful to Leonardo Levinas for
explaining this important point to me.}.

In this sense we will argue it is wrong to follow the orthodox
reductionistic conception: {\it  that everything is ``quantum" and
that it is necessary to find a limit-process between quantum
mechanics and our classical conception of the world (the quantum
to classical limit)}; rather one should be able to acknowledge the
possibility of describing the world from different (incompatible
but complementary) view points. Wolfgang Pauli imagined a future
in which the quantum conceptions would not be regarded as `weird.'
This conceptual jump in the way we see the world has not yet taken
place with respect to quantum mechanics\footnote{See also
Constantin Piron \cite{Piron}.}. We believe this is due to the
lack of new concepts in the quantum domain and to a
`reductionistic conception' which presupposes that `understanding'
is reducible to `classical understanding', to a single view point.
My aim in \cite{deRondeCDII} will be to show that quantum
mechanics is {\it not} an open theory, in the sense of
convergence\footnote{It is important to point out that the view of
Dirac is read as implying convergence of theories.}, that it is
not necessary to find a reduction between classical mechanics and
quantum mechanics. Rather, that {\it the quantum and classical
descriptions are complementary} with respect to the concepts
involved, thus, that quantum mechanics, nor classical mechanics,
can be regarded as a fundamental theories in this reductionistic
sense. The {\it classical description} is complementary to the
{\it quantum description} and even though they can make contact in
some definite situations, one should acknowledge the structural
and conceptual impossibility of the quantum theory to explain the
classical world (this will be further analyzed in
\cite{deRondeCDII}, Sect. 1).

Scientific realism is the position that theory construction aims
to give us a literally true story of what the world is like, and
that acceptance of a scientific theory involves the belief that it
is true. The idea of truth as a closed enterprize is responsible
to great extent for the development of the `fabric of science'.
Reductionism allows a single truth, it is hostile to every
conception which is outside its own limits. But truth is nothing
but a concept created by man. Max Planck expressed this point in a
very bright way: {\it ``A new scientific truth does not triumph by
convincing its opponents and making them see the light, but rather
because its opponents eventually die, and a new generation grows
up that is familiar with it."} Quantum mechanics presents us a new
truth, with new concepts which up to the present have not been
further developed. In \cite{deRondeCDII} I will present a new way
of `reading' quantum mechanics through the distinction of
different {\it complementary levels of description}. I will
discriminate between different levels in order to clarify the
discussion and show that different, mutually incompatible concepts
are taken into account in the discussion of the interpretational
problems of the quantum theory.

\section*{Conclusions}

I have presented a development of the concept of complementarity
following the road shown (specially) by Niels Bohr, Werner
Heisenberg and Wolfgang Pauli. I see this as a starting point in
the development of the interpretation of the quantum theory and a
return to the philosophical attitude which was brought by the
founding fathers of the quantum theory. To return we must take
into account the very deep philosophical understanding of the
time, and go still further in this forgotten path.

\section*{Acknowledgments}

I specially want to thank Diederik Aerts, Dennis Dieks and
Graciela Domenech for encouragement and guidance on my work. I am
indebt with Sven Aerts, Karin Verelst, Sonja Smets, Soazig Le
Bihan, Michiel Seevinck and Patricia Kauark for stimulating
discussions and comments on earlier drafts of this article.


\begin{thebibliography}{99}

\bibitem{Aerts85} Aerts, D., 1985, ``The physical origin of the EPR paradox and how
to violate Bell inequalities by macroscopical systems" In {\it On
the Foundations of Modern Physics}, P. Lathi, and P. Mittelstaedt
(eds.) , World Scientific, Singapore, pp.305-320.

\bibitem{Aerts88} Aerts, D., 1988, ``The entity and modern physics: the
creation-discovery view of reality" In {\it Interpreting Bodies:
Classical and Quantum Objects in Modern Physics}, E. Castellani
(ed.), Princeton University Press, Princeton.

\bibitem{AertsdeRonde} Aerts, D., de Ronde, C. and D'Hooghe B., 2005,
``Compatibility and Separability for Classical and Quantum
Entanglement" submitted to {\it International Journal of
Theoretical Physics}.

\bibitem{BeneDieks02} Bene, G. and Dieks, D., 2002, ``A Perspectival Version of
the Modal Interpretation of Quantum Mechanics and the Origin of
Macroscopic Behavior", {\it Foundations of Physics}, {\bf 32}, No
5, May 2002, also in archive ref and link: quant-ph/0112134.

\bibitem{Bohr28} Bohr, N., 1928, ``The Quantum Postulate and the recent
development of atomic theory", {\it Nature}, {\bf 121},
pp.580-590.

\bibitem{Bohr58} Bohr, N., 1958, {\it Atomic, Physics and Human Knowledge}, Wiley, New York.

\bibitem{Bokulich04} Bokulich, A., 2004, ``Open or Closed? Dirac,
Heisenberg, and the relation between classical and quantum
mechanics", {\it Studies in History and Phylosophy of Modern
Physics}, {\bf 35}, pp.377-396.

\bibitem{Cassirer56} Cassirer, E., 1956, {\it Determinism and
Indeterminism in Modern Physics}, Yale University Press.

\bibitem{D'Espagnat95} D'Espagnat, B., 1995, {\it Veiled Reality. An Analysis
of Present Day Quantum Mechanical Concepts}, Addison-Wesley,
Reading MA.

\bibitem{Dieks88a} Dieks, D., 1988, ``The Formalism of Quantum Theory: An
Objective description of reality", {\it Annalen der Physik}, {\bf
7}, Band 45, Heft 3, pp.174-190.

\bibitem{Feyerabend93} Feyerabend, P., 1993, {\it Against Method}, Third Edition, London: Verso.

\bibitem{Finkelstein87} Finkelstein, D., 1987, ``All is flux" In
{\it Quantum Implications: Essays in honour of David Bohm},
pp.289-94, B.J. Hiley and F.D. Peat (eds.), London: Routledge and
Kegan Paul.

\bibitem{Folse85} Folse, H.J., 1985, {\it The Philosophy of Niels Bohr:
The Framework of Complementarity}, North Holland Physics
Publishing, Amsterdam.

\bibitem{Folse87} Folse, H.J., 1987, ``Niels Bohr's Concept of Reality",
In {\it Symposium on the foundations of Modern Physics 1987},
pp.161-179, P. Lathi and P. Mittelstaedt (eds.), World Scientific,
Singapore.

\bibitem{Healey99} Healey, R.A., 1999, ``Holism and Non Separability in
Physics", The Stanford Encyclopedia of Philosophy (Winter 2004
Edition), Edward N. Zalta (ed.), URL =
http://plato.stanford.edu/archives/win2004/entries/physics-holism/.

\bibitem{Heisenberg27} Heisenberg, W., 1927, ``Uber den anschaulichen
Inhalt der quantentheoretischen Kinematik und Mechanic" {\it
Zeitschrift fur Physik}, {\bf 43}, pp.172-98; reprinted as ``The
Physical Content of Quantum Kinematics and Mechanics", translation
by J.A. Wheeler and W.H. Zurek, in {\it Quantum Theory and
Measurement}, J.A. Wheeler and W.H. Zurek (eds.).

\bibitem{Heisenberg72} Heisenberg, W., 1972, {\it Dialogos sobre la F\'isica At\'omica},
Biblioteca de Autores Cristianos de la Editorial Cat\'olica,
Madrid.

\bibitem{Hendry84} Hendry, J., 1984, {\it The Creation of Quantum
Mechanics and the Bohr-Pauli Dialogue}, D. Reidel Publishing
Company, Dordrecht.

\bibitem{Howard89} Howard, D., 1989, ``Holism, Separability and
the Metaphysical implications of the Bell inequalities", In {\it
Philosophical Consequences of Quantum Theory: Reflections on
Bell's Theorem}, pp.224-253, Cushing and McMullin (eds.),
University of Notre Dame Press, Notre Dame, Indiana.

\bibitem{Howard94} Howard, D., 1994, ``What makes a classical
concept classical? Towards a reconstruction of Niels Bohr's
philosophy of physics" In {\it Niels Bohr and Contemporary
Philosophy}, J. Faye and H. Folse (eds.), Kluwer, Boston.

\bibitem{Jammer74} Jammer, M., 1974, {\it The Philosophy of Quantum
Mechanics}, Wiley, New York.

\bibitem{Kant} Kant, I., 1973, {\it Critique of Pure Reason}, Trans. Norman Kemp Smith. London: Macmillan.

\bibitem{Laurikainen98} Laurikainen, K.V., 1998, {\it The
Message of the Atoms, Essays on Wolfgang Pauli and the
Unspeakable}, Spinger Verlag, Berlin.

\bibitem{Lombardi02} Lombardi, O., 2002, ``Determinism, Internalism and
Objectivity", In {\it Between Chance and Choice}, pp.75-87, H.
Atmanspacher and R. Bishop (eds.), Imprint Academic, Exeter.

\bibitem{Pauli94} Pauli, W., 1994, {\it Writings on Physics and
Philosophy}, Enz, C. and von Meyenn, K. (eds.), Springer Verlag.

\bibitem{PauliJung} Pauli, W. and Jung, C.G., 2001, {\it Atom and
Archetype, The Pauli/Jung Letters 1932-1958}, Princeton University
Press, New Jersey.

\bibitem{Petersen63} Petersen, A., 1963, ``The philosophy of Niels
Bohr", {\it The Bulletin of the Atomic Scientists}, September
1963.

\bibitem{Piron} Piron, C., 1999, ``Quanta and Relativity: Two Failed Revolutions",
In {\it The White Book of Einstein Meets Magritte}, pp.107-112, D.
Aerts J. Broekaert and E. Mathijs (eds.), Kluwer Academic
Publishers.

\bibitem{Plotnitsky94} Plotnitsky, A., 1994, {\it Complementarity}, Duke University Press, Durham and London.

\bibitem{Primas83} Primas, H., 1983, {\it Chemistry, Quantum Mechanics and
Reductionism}, Springer Verlag, Berlin.

\bibitem{Primas87} Primas, H. 1987, ``Contextual Quantum Objects and
their Ontic Interpretation", In {\it Symposium on the foundations
of Modern Physics 1987}, pp.251-275, P.Lathi and P. Mittelstaedt
(eds.), World Scientific, Singapore.

\bibitem{Primas94} Primas, H., 1994, ``Hierarchical Quantum Descriptions
and their Associated Ontologies" In {\it Symposia on the
Foundations of Modern Physics 1994}, pp.201-220, Laurikainen,
K.V., Montonen C. and Sunnarborg K. (eds.), Frontiers,
Gif-sur-Yvette Cedex.

\bibitem{Putnam81} Putnam, H., 1981, {\it Reason, Truth and History}, Cambridge
University Press, Cambridge.

\bibitem{deRonde03} de Ronde, C., 2003, Master Thesis: {\it Perspectival
Interpretation of Quantum Mechanics (a story about correlations
and holism)}, Institute for History and Foundations of
Mathematical and the Natural Sciences, Utrecht University and
University of Buenos Aires, URL =
http://www.vub.ac.be/CLEA/people/deronde/.

\bibitem{deRonde04} de Ronde, C., 2004, ``Interpretaci\'on
perspectival de la mec\'anica cu\'antica y descripciones
complementarias" In {\it Volumen 10 de Epistemolog\'ia e Historia
de la Ciencia}, pp.161-167, Garcia, P. and Morey, P. (eds.),
Universidad Nacional de Cordoba, Cordoba. URL =
http://www.vub.ac.be/CLEA/people/deronde/.

\bibitem{deRondeCDII} de Ronde, C., 2005, ``Complemenary Descriptions
(PART II): A Set of Ideas Regarding the Interpretation of Quantum
Mechanics", Preprint.

\bibitem{Rovelli96} Rovelli, C., 1996, ``Relational Quantum Mechanics", archive ref and link: quant-ph/9609002.

\bibitem{Saunders05} Saunders, S., 2005, ``Complementarity
and Scientific Rationality", {\it Foundations of Physics},
forthcoming.

\bibitem{Schrodinger35} Schr\"odinger, E., 1935, ``The Present
Situation in Quantum Mechanics", {\it Naturwiss}, {\bf 23}, 807,
translated to english in {\it Quantum Theory and Measurement},
J.A. Wheeler and W.H. Zurek (eds.), Princeton University Press
1983.

\bibitem{VerelstCoecke99} Verelst, K. and Coecke, B., 1999,
``Early Greek Thought and perspectives for the Interpretation of
Quantum Mechanics: Preliminaries to an Ontological Approach" In
{\it The Blue Book of Einstein Meets Magritte}, pp.163-196, D.
Aerts (eds.), Kluwer Academic Publishers.

\bibitem{vWeizsacker55} Von Weizs\"acker, C.F., 1955, ``Komplementarit\"at
und Natuurwissenschaft", {\it Die Natuurwissenschaften}, p.42,
n.19-20. Translated as ``Complementariedad y L\'ogica" in {\it La
Imagen F\'isica del Mundo}, 1974, Biblioteca de Autores
Cristianos, Madrid.

\bibitem{vWeizsacker74} Von Weizs\"acker, C.F., 1974, {\it La Imagen F\'isica del Mundo}, Biblioteca de Autores
Cristianos, Madrid.


\bibitem{vWeizsacker85} Von Weizs\"acker, C.F., 1985,
``Heisenberg's philosophy" In {\it Symposium on the Foundations of
Modern Physics 1985}, pp.277-293, P. Lathi and P. Mittelstaedt
(eds.), World Scientific, Singapore.

\bibitem{Weinberg93} Weinberg, 1993, {\it Dreams of a final theory},
Vintage, London.

\bibitem{WheelerZurek83} Wheeler, J.A. and Zurek, W.H., 1983, {\it Quantum Theory and
Measurement}, J.A. Wheeler and W.H. Zurek (eds.), Princeton
University Press, New Jersey.

\bibitem{Wittgenstein} Wittgenstein, L., 1974, {\it Tractatus Logico
Philosophicus}, Routledge Classics, London.

\end{thebibliography}
\end{document}